\documentstyle[12pt,a41,axodraw,epsfig]{article}
\begin{document}
\mbox{}
\begin{center}
{\large\bf Production of Heavy Quarks in Deep-Inelastic Lepton-Hadron 
Scattering\footnote{Talk presented at the Migneron Tribute Session of the 
conference MRST 2001, The University of Western Ontario, London, Canada, 
May 15-18}}\\
\vspace*{20mm}
\large
{W.L.van Neerven\footnote{Work supported by the 
EC network `QCD and Particle Structure'  under contract No.~FMRX--CT98--0194.}}
\\
\vspace{2em}
\normalsize
{Instituut-Lorentz, Universiteit Leiden, P.O. Box 9506, 2300 RA Leiden,
The Netherlands.}
\end{center}
\vspace*{10mm}
\begin{abstract}
We will give a review of the computation of exact next-to-leading order
corrections to heavy quark production in deep inelastic lepton-hadron
scattering and discuss the progress made in this field over the past
ten years. In this approach, hereafter called ${\rm EXACT}$, where the heavy 
quark mass is taken to be of the same order of magnitude as the other large 
scales in the process, one can apply perturbation theory in all orders of
the strong coupling constant $\alpha_s$. The results are compared with
another approach, called the {\it variable flavor number scheme} 
(${\rm VFNS}$), where the heavy quark is also treated as a massless quark. 
It turns out that the differences between the two approaches are very small 
provided both of them are carried out up to next-to-next-to-leading order. 
\end{abstract}
\begin{center}
\section{Introduction}
\end{center}

In the last ten years one has made much progress on the theoretical and
experimental level in the study of heavy flavor production in deep inelastic 
lepton-hadron scattering. Computations of the cross section in the Born
approximation, where the heavy quark is treated as a massive particle, were 
finished by the end of the seventies \cite{wit} 
whereas the first experimental results came from the EMC-collaboration 
\cite{aubert} in the early eighties. In the nineties the ${\cal O}(\alpha_s)$
corrections to the Born process were computed in \cite{lrsn}. Moreover
other methods to study heavy flavor production were advocated like the
intrinsic quark approach \cite{bhmt} and the variable flavor number scheme 
\cite{acot} where the heavy quark is also treated as a massless particle.
From the experimental side the electron (positron)-proton collider HERA
has given us a wealth of information about charm quark production (for recent
experimental results see \cite{breit}, \cite{adlof}, \cite{zeus}, \cite{H1}).
We will report about this progress below.

\begin{figure}
\begin{center}
  \begin{picture}(190,140)(0,0)
    \ArrowLine(0,20)(80,30)
    \ArrowLine(80,40)(160,80)
    \ArrowLine(80,30)(157,35)
    \ArrowLine(80,25)(170,25)
    \ArrowLine(80,20)(157,15)
    \Line(150,40)(170,25)
    \Line(150,10)(170,25)
    \GCirc(80,30){20}{0.3}
    \Photon(50,100)(80,40){3}{7}
    \DashArrowLine(0,100)(50,100){5}
    \DashArrowLine(50,100)(100,120){5}
    \Text(0,113)[t]{$e^-$}
    \Text(100,133)[t]{$e^-$}
    \Text(0,15)[t]{$P$}
    \Text(170,85)[t]{$Q$}
    \Text(52,80)[t]{$V$}
    \Text(180,30)[t]{$'X'$}
    \Text(40,20)[t]{$p$}
    \Text(25,115)[t]{$l_1$}
    \Text(75,125)[t]{$l_2$}
    \Text(75,80)[t]{$\downarrow q$}
    \Text(120,55)[t]{$p_1$}
  \end{picture}
  \caption[]{Kinematics of heavy quark $Q$-production in deep inelastic
             electron-proton scattering}
  \label{fig:1}
\end{center}
\end{figure}
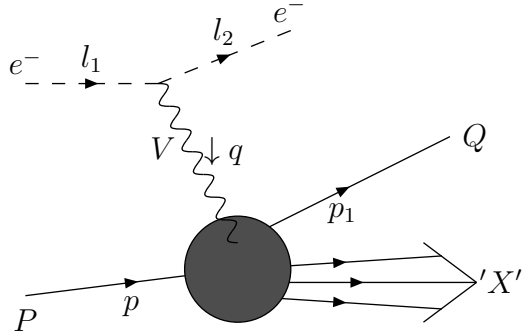

The semi-inclusive process describing heavy quark production in deep
inelastic electron-proton scattering is given by (see Fig. \ref{fig:1})
\begin{eqnarray}
\label{I1}
e^-(k_1)+P(p) \rightarrow e^-(k_2)+Q(p_1) + 'X'\,,
\end{eqnarray}
where $V$ represents the intermediate vector boson $Z$ or $\gamma$
carrying the momentum $q$ and $Q$ denotes the heavy (anti-) quark. 
The scaling variables are given by
\begin{eqnarray}
\label{I2}
&& x=\frac{Q^2}{2p\cdot q}\,, \qquad y=\frac{p\cdot q}{p\cdot k_1}\,,
\nonumber\\[2ex]
q^2=(k_1-k_2)^2 &\equiv& -Q^2 <0\,, \qquad 0<x\le 1\,, \qquad 0<y< 1\,.
\end{eqnarray}
In the ongoing experiments carried out at HERA \cite{breit}, \cite{adlof}, 
\cite{zeus}, \cite{H1} and in the fixed target experiments carried out in 
the past \cite{aubert}, $Q^2 \ll M_Z^2$ so that the intermediate
$Z$-boson can be neglected and the reaction is dominated by photon
exchange. In this case the inclusive cross section
simplifies considerably and when the incoming particles are unpolarised
it can be written as
\begin{eqnarray}
\label{I3}
\frac{d^2\sigma}{dx\,dQ^2}=\frac{2\pi\alpha^2}{xQ^4}\Bigg [
\Big \{1 + (1-y)^2\Big \}\,F_{2,Q}(x,Q^2,m^2)-y^2\,F_{L,Q}(x,Q^2,m^2)\Bigg ]
\,,
\end{eqnarray}
where $F_{k,Q}$ ($k=2,L$) denote the heavy quark contributions to the
structure functions. Notice that an analogous formula exists for
the semi-inclusive cross section
\begin{eqnarray}
\label{I4}
\frac{d^4\sigma}{dx\,dy\,dp_{T,Q}\,d\eta_Q}\,,
\end{eqnarray}
where $p_{T,Q}$ and $\eta_Q$ denote the transverse momentum and rapidity
of the heavy quark $Q$ respectively. Both of them are considered in the
center of mass of the photon-proton system.
For neutral current reactions one has proposed two different production
mechanisms for heavy quark production. They are distinguished as follows
\begin{itemize}
\item[I]
\underline{Intrinsic Heavy Quark Production}\\[3mm]
Here one assumes that, besides light quarks $u,d,s$ and gluons $g$,
the wave function of the proton also consists of heavy quarks like
$c,b,t$ \cite{bhmt}. In the context of the QCD improved parton model this 
means that
the production mechanism is described as indicated in Fig. \ref{fig:2}.
In this picture the heavy quark emerges directly from the proton and
interacts with the virtual photon $\gamma^*$. The consequence is that
it is described by a heavy flavor density $f_Q(z,\mu^2)$ with $p_Q=z\,p$
and $\mu$ denotes the factorization scale. For this mechanism the heavy
quark structure function has the following representation
\begin{eqnarray}
\label{I5}
F_{k,Q}(x,Q^2,m^2)&=&e_Q^2\,\int_x^1\frac{dz}{z}\,f_Q(x/z,\mu^2)\,
H_{k,Q}(z,Q^2,m^2,\mu^2)
\nonumber\\[2ex]
&\equiv& e_Q^2\,f_Q(\mu^2)\otimes H_{k,Q}(Q^2,m^2,\mu^2)\,.
\end{eqnarray}
Here $e_Q$ denotes the charge of the heavy quark and $\otimes$ stands for 
the convolution
\begin{eqnarray}
\label{I6}
f \otimes g(x)=\int_x^1\frac{dz}{z}\,f(x/z)\,g(z)\,.
\end{eqnarray}
Further the heavy quark coefficient function $H_{k,Q}$ can be expanded
in the strong coupling constant $\alpha_s(\mu^2)$ as follows
\begin{eqnarray}
\label{I7}
H_{k,Q}=\sum_{n=0}^{\infty}a_s^n\,H_{k,Q}^{(n)}\,,\quad \mbox{with}
\quad a_s\equiv \frac{\alpha_s}{4\pi}\,.
\end{eqnarray}
Some of the contributions to $H_{k,Q}$ are given by the diagrams in
Fig. \ref{fig:2}a,b.
\item[II]
\underline{Extrinsic Heavy Quark Production}\\[3mm]
In this case the proton wave function does not contain the heavy quark
components. In lowest order of perturbation theory the heavy quark and
heavy anti-quark appear in pairs and are produced via photon-gluon
fusion \cite{wit} as presented in Fig. \ref{fig:3}. Here the gluon emerges 
from the proton in a similar way as the heavy quark in Fig. \ref{fig:2}.
In this approach the heavy quark structure function reads in lowest order
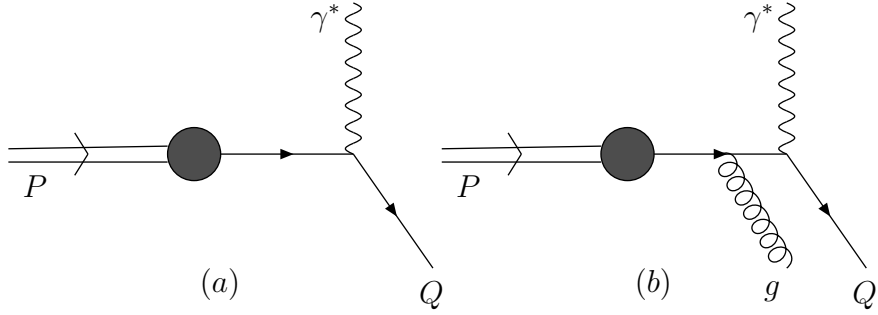
\begin{figure}
\begin{center}
  \begin{picture}(160,150)(0,0)
    \Line(25,71)(30,63)
    \Line(25,55)(30,63)
    \Line(0,60)(60,60)
    \Line(0,65)(60,66)
    \GCirc(70,63){10}{0.3}
    \Photon(130,120)(130,63){3}{7}
    \ArrowLine(80,63)(130,63)
    \ArrowLine(130,63)(160,20)
    \Text(120,120)[t]{$\gamma^*$}
    \Text(160,15)[t]{$Q$}
    \Text(10,55)[t]{$P$}
    \Text(80,20)[t]{$(a)$}
  \end{picture}
  \begin{picture}(160,150)(0,0)
    \Line(25,71)(30,63)
    \Line(25,55)(30,63)
    \Line(0,60)(60,60)
    \Line(0,65)(60,66)
    \GCirc(70,63){10}{0.3}
    \Photon(130,120)(130,63){3}{7}
    \Gluon(105,63)(130,20){4}{7}
    \ArrowLine(80,63)(130,63)
    \ArrowLine(130,63)(160,20)
    \Text(125,15)[t]{$g$}
    \Text(120,120)[t]{$\gamma^*$}
    \Text(160,15)[t]{$Q$}
    \Text(10,55)[t]{$P$}
    \Text(80,20)[t]{$(b)$}
  \end{picture}
  \caption[]{$(a)$ $\gamma^* + Q \rightarrow Q$ ($H_{k,Q}^{(0)}$), $(b)$ 
$\gamma^* + Q \rightarrow Q + g$ ($H_{k,Q}^{(1)}$).}
  \label{fig:2}
\end{center}
\end{figure}
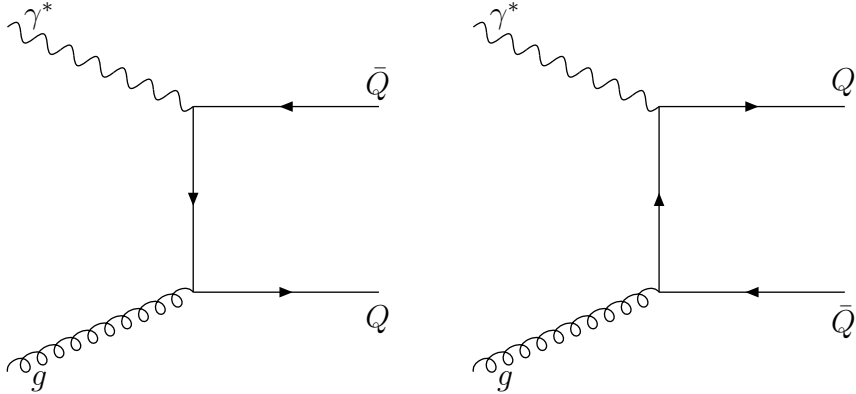
\begin{figure}
\begin{center}
  \begin{picture}(140,140)
  \Gluon(0,0)(70,30){3}{10}
  \ArrowLine(140,100)(70,100)
  \ArrowLine(70,100)(70,30)
  \ArrowLine(70,30)(140,30)
  \Photon(0,130)(70,100){3}{7}
    \Text(12,0)[t]{$g$}
    \Text(12,140)[t]{$\gamma^*$}
    \Text(140,115)[t]{$\bar Q$}
    \Text(140,25)[t]{$Q$}
  \end{picture}
\hspace*{1cm}
  \begin{picture}(140,140)
  \Gluon(0,0)(70,30){3}{10}
  \ArrowLine(70,100)(140,100)
  \ArrowLine(70,30)(70,100)
  \ArrowLine(140,30)(70,30)
  \Photon(0,130)(70,100){3}{7}
    \Text(12,0)[t]{$g$}
    \Text(12,140)[t]{$\gamma^*$}
    \Text(140,115)[t]{$Q$}
    \Text(140,25)[t]{$\bar Q$}
  \end{picture}
\vspace*{5mm}
  \caption{Feynman diagrams for the lowest-order photon-gluon
           fusion process contributing to the coefficient functions
           $H_{k,g}^{(1)}$.}
  \label{fig:3}
\end{center}
\end{figure}
\begin{eqnarray}
\label{I8}
F_{k,Q}(x,Q^2,m^2)= a_s\,e_Q^2\,f_g(\mu^2)\otimes 
H_{k,g}^{(1)}(Q^2,m^2,\mu^2)\,.
\end{eqnarray}
\end{itemize}
The main difference between the two production mechanisms can be attributed
to the fact
that for extrinsic heavy quark production two heavy particles are produced in 
the final state instead of one as in the case of the intrinsic heavy quark
approach. This reveals itself in the transverse momentum $p_T$-distribution
where for mechanism ${\rm II}$ the quark and anti-quark appear back to back.
The experiments carried out HERA \cite{H1} confirm the $p_T$-spectrum 
predicted by the latter mechanism. However in the past the EMC-collaboration 
\cite{aubert} found a discrepancy at large $x$ ($x=0.237$) between the 
structure function $F_{2,Q}$, predicted by mechanism ${\rm II}$, and the charm 
quark data. This difference can be explained by also invoking mechanism 
${\rm I}$ \cite{bhmt} (see also \cite{hsv}). Nevertheless because
of the success of extrinsic heavy quark production revealed by the HERA 
charm quark data we will continue with this approach in our calculations 
below.
\begin{center}
\section{Heavy Quark Structure Functions up to ${\cal O}(\alpha_s^2)$}
\end{center}

In the extrinsic heavy quark approach there are two different production 
mechanisms \cite{bmsmn}, \cite{neer1} which are of importance for the derivation
of the variable flavor number scheme (VFNS) treated in the next section.
In the case of electro-production the virtual photon either interacts 
with the heavy quark appearing in the final state or it is attached to the 
light (anti-) quark in the initial state. This distinction is revealed by 
the Feynman graphs. Some of them will be shown as illustration below.
\begin{itemize}
\item[A]
\underline{The photon interacts with the heavy quark}\\[3mm]
Here the lowest order (LO) process is given by gluon fusion as presented in
Fig. \ref{fig:3}. It is given by the partonic reaction
\begin{eqnarray}
\label{H1}
\gamma^* + g \rightarrow Q + \bar Q\,,
\end{eqnarray}
which is calculated in \cite{wit}.
In next-to-leading order (NLO) one has to compute the virtual corrections to 
this process. Some of the Feynman graphs are shown in 
Fig. \ref{fig:4}.
\begin{figure}
\begin{center}
  \begin{picture}(60,60)
  \Gluon(0,0)(30,15){3}{7}
  \ArrowLine(60,15)(30,15)
  \ArrowLine(30,15)(30,45)
  \ArrowLine(30,45)(60,45)
  \Gluon(45,45)(45,15){3}{7}
  \Photon(0,60)(30,45){3}{7}
  \end{picture}
\hspace*{1cm}
  \begin{picture}(60,60)
  \Gluon(0,0)(30,15){3}{7}
  \ArrowLine(60,15)(30,15)
  \ArrowLine(30,15)(30,45)
  \ArrowLine(30,45)(60,45)
  \Gluon(30,30)(45,15){3}{7}
  \Photon(0,60)(30,45){3}{7}
\end{picture}
\hspace*{1cm}
  \begin{picture}(60,60)
  \Gluon(0,0)(30,15){3}{7}
  \ArrowLine(60,15)(30,15)
  \ArrowLine(30,15)(30,45)
  \ArrowLine(30,45)(60,45)
  \Gluon(30,30)(45,45){3}{7}
  \Photon(0,60)(30,45){3}{7}
\end{picture}
\hspace*{1cm}
  \begin{picture}(60,60)
  \Gluon(0,0)(30,15){3}{7}
  \ArrowLine(60,15)(30,15)
  \ArrowLine(30,15)(30,45)
  \ArrowLine(30,45)(60,45)
  \Gluon(15,7)(45,45){3}{7}
  \Photon(0,60)(30,45){3}{7}
    \Text(9,0)[t]{$k_1$}
    \Text(45,35)[t]{$k$}
\end{picture}
\vspace*{5mm}
  \caption{Virtual gluon corrections to the process $\gamma^* + g \rightarrow
           Q + \bar Q$ contributing to the coefficient functions
           $H_{i,g}^{(2)}$.}
  \label{fig:4}
\end{center}
\end{figure}
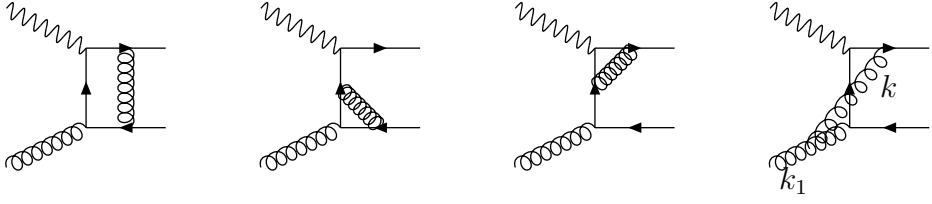
Besides the virtual corrections one also has to calculate all processes which
contain an extra particle in the final state. Adding a gluon to reaction
(\ref{H1}) we observe gluon bremsstrahlung given by
\begin{eqnarray}
\label{H2}
\gamma^* + g \rightarrow Q + \bar Q + g\,.
\end{eqnarray}
Some of the Feynman graphs corresponding to the process above are presented in
Fig. \ref{fig:5}. Besides this reaction we have another one which appears
for the first time if the computations are extended to NLO. It is represented
by the Bethe-Heitler process 
\begin{eqnarray}
\label{H3}
\gamma^* + q(\bar q) \rightarrow Q + \bar Q + q(\bar q)\,,
\end{eqnarray}
where the Feynman diagrams are shown in Fig. \ref{fig:6}.
\begin{figure}
\begin{center}
  \begin{picture}(60,60)
  \Gluon(30,0)(30,30){3}{7}
  \ArrowLine(60,30)(30,30)
  \ArrowLine(30,30)(30,45)
  \ArrowLine(30,45)(60,45)
  \Gluon(30,15)(60,15){3}{7}
  \Photon(0,60)(30,45){3}{7}
    \Text(20,10)[t]{$k_1$}
    \Text(55,10)[t]{$k_2$}
  \end{picture}
\hspace*{1cm}
  \begin{picture}(60,60)
  \Gluon(30,0)(30,15){3}{7}
  \ArrowLine(60,15)(30,15)
  \ArrowLine(30,15)(30,45)
  \ArrowLine(30,45)(60,45)
  \Gluon(30,30)(60,30){3}{7}
  \Photon(0,60)(30,45){3}{7}
\end{picture}
\hspace*{1cm}
  \begin{picture}(60,60)
  \Gluon(30,0)(30,15){3}{7}
  \ArrowLine(60,15)(30,15)
  \ArrowLine(30,15)(30,45)
  \ArrowLine(30,45)(60,45)
  \Gluon(45,15)(60,0){3}{7}
  \Photon(0,60)(30,45){3}{7}
\end{picture}
\hspace*{1cm}
  \begin{picture}(60,60)
  \Gluon(0,0)(30,15){3}{7}
  \ArrowLine(60,15)(30,15)
  \ArrowLine(30,15)(30,45)
  \ArrowLine(30,45)(60,45)
  \Gluon(45,45)(60,60){3}{7}
  \Photon(0,60)(30,45){3}{7}
\end{picture}
\vspace*{5mm}
  \caption{The  bremsstrahlung process $\gamma^* + g \rightarrow
           Q + \bar Q + g$ contributing to the coefficient functions
           $H_{i,g}^{(2)}$.}
  \label{fig:5}
\end{center}
\end{figure}
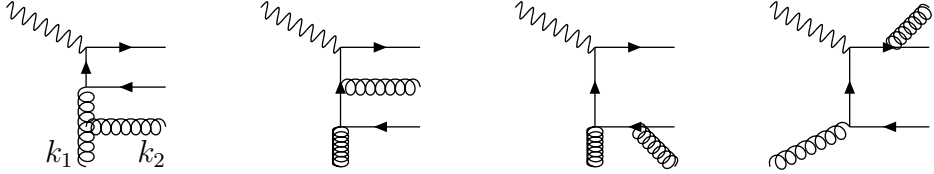
The contributions to the heavy quark
structure functions due to the reactions above can be written as 
\begin{eqnarray}
\label{H4}
F_{k,Q}(Q^2,m^2)&=&e_Q^2\Bigg [f_g(3,\mu^2)\otimes \Big \{a_s(\mu^2)
H_{k,g}^{(1)}(Q^2,m^2)+a_s^2(\mu^2)H_{k,g}^{(2)}(Q^2,m^2,\mu^2)\Big \}
\nonumber\\[2ex]
&& +a_s^2(\mu^2)f_q^{\rm S}(3,\mu^2)\otimes H_{k,q}^{(2)}(Q^2,m^2,\mu^2)\Bigg ]
\,.
\end{eqnarray}
\begin{figure}
\begin{center}
  \begin{picture}(60,60)
  \DashArrowLine(0,0)(30,0){3}
  \DashArrowLine(30,0)(60,0){3}
  \Gluon(30,0)(30,25){3}{7}
  \ArrowLine(60,25)(30,25)
  \ArrowLine(30,25)(30,50)
  \ArrowLine(30,50)(60,50)
  \Photon(0,60)(30,50){3}{7}
  \end{picture}
\hspace*{1cm}
  \begin{picture}(60,60)
  \DashArrowLine(0,0)(30,0){3}
  \DashArrowLine(30,0)(60,0){3}
  \Gluon(30,0)(30,25){3}{7}
  \ArrowLine(30,25)(60,25)
  \ArrowLine(30,50)(30,25)
  \ArrowLine(60,50)(30,50)
  \Photon(0,60)(30,50){3}{7}
\end{picture}
\vspace*{5mm}
  \caption{The Bethe-Heitler process $\gamma^* + Q \rightarrow
           Q + \bar Q + q$ contributing to the coefficient functions
           $H_{i,q}^{(2)}$.}
  \label{fig:6}
\end{center}
\end{figure}
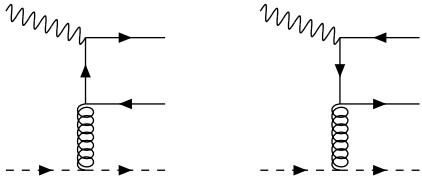
Here $e_Q$ denotes the charge of $Q$ indicating that the photon couples 
to the heavy quark and $H_{k,i}$ ($i=q,g$), characteristic of these type of 
processes, represent the heavy quark coefficient functions
which emerge from the calculation of the Feynman graphs after renormalization
and mass factorization is carried out. The former procedure leads to a 
dependence on a renormalization scale $\mu$ of the coupling constant $a_s$, the 
coefficient function $H_{k,i}$ and the parton densities $f_i$. Moreover the 
latter two also depend on a factorization scale which for convenience
is set equal to the parameter $\mu$ defined above. The factorization scale,
which can be attributed to mass factorization, appears in all coefficient 
functions except for the lowest order one i.e. $H_{k,i}^{(1)}$. In the
structure function of Eq. (\ref{H4}) there appear two different types of 
flavor singlet parton densities i.e. the gluon density $f_g$ and the quark
singlet density which in a three flavor number scheme reads as
\begin{eqnarray}
\label{H5}
&& f_q^{\rm S}(3,\mu^2)=
\nonumber\\[2ex]
&&f_u(3,\mu^2)+f_{\bar u}(3,\mu^2)+f_d(3,\mu^2)
+f_{\bar d}(3,\mu^2)+f_s(3,\mu^2)+f_{\bar s}(3,\mu^2)\,.
\end{eqnarray}
Hence Eq. (\ref{H4}) represents the singlet part of the heavy quark structure
function.
\item[B]
\underline{The photon interacts with the light quark}.\\[3mm]
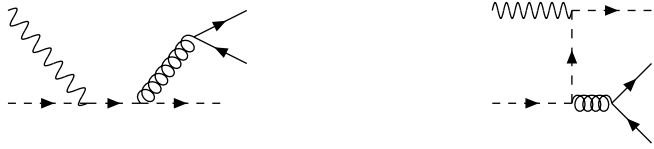
\begin{figure}
\begin{center}
  \begin{picture}(90,50)
  \DashArrowLine(0,15)(30,15){3}
  \DashArrowLine(30,15)(50,15){3}
  \DashArrowLine(50,15)(80,15){3}
  \ArrowLine(70,40)(90,50)
  \ArrowLine(90,30)(70,40)
  \Gluon(50,15)(70,40){3}{7}
  \Photon(0,50)(30,15){3}{7}
  \end{picture}
\hspace*{3cm}
  \begin{picture}(60,50)
  \DashArrowLine(0,15)(30,15){3}
  \DashArrowLine(30,15)(30,50){3}
  \DashArrowLine(30,50)(60,50){3}
  \Gluon(30,15)(45,15){3}{4}
  \ArrowLine(45,15)(60,30)
  \ArrowLine(60,0)(45,15)
  \Photon(0,50)(30,50){3}{7}
\end{picture}
\vspace*{5mm}
  \caption{The Compton process $\gamma^* + Q \rightarrow
           Q + \bar Q + q$ contributing to the coefficient functions
           $L_{i,q}^{(2)}$.}
  \label{fig:7}
\end{center}
\end{figure}
The second production mechanism is represented by the Feynman diagrams
where the photon couples to the light quark or the anti-quark. This happens
for the first time in NLO where one observes the Compton process
\begin{eqnarray}
\label{H6}
\gamma^* + q(\bar q) \rightarrow Q + \bar Q + q(\bar q)\,,
\end{eqnarray}
which is depicted in Fig. \ref{fig:7}. The coefficient functions
corresponding to this type of process are denoted by $L_{k,i}$ ($i=q,g$)
and the contribution to the heavy quark structure functions is characterized
by the expression
\begin{eqnarray}
\label{H7}
F_{k,Q}(Q^2,m^2)&=&\sum_{i=u,d,s}e_i^2\,a_s^2(\mu^2)\,\Big (f_i(3,\mu^2)
+f_{\bar i}(3,\mu^2)\Big )\otimes L_{k,q}^{(2)}(Q^2,m^2)\,,
\end{eqnarray}
where $e_i$ denotes the charge of the light quark represented by $i=u,d,s$
in a three flavor number scheme. Since this process appears for the first
time in second order mass factorization is not needed which explains
the independence of $L_{k,q}^{(2)}$ on the parameter $\mu$.
\end{itemize}
The computation of the second order contributions to the heavy quark
coefficient functions $H_{k,i}$, $L_{k,i}$ has been carried out in \cite{lrsn}.
While calculating the Feynman graphs in Figs. \ref{fig:4}-
\ref{fig:6} one encounters several type of singularities which have to
be regularized and subsequently to be subtracted off before one obtains a 
finite result. The singularities are of the following nature i.e. infrared (IR),
ultraviolet (UV) and collinear (C). Sometimes the latter are also called
mass singularities. The IR divergences cancel between the virtual and
the bremsstrahlung corrections to reaction (\ref{H1}). The UV divergences,
regularized by n-dimensional regularization, are removed by mass
and coupling constant renormalization. For the mass renormalization we
choose the on-shell scheme. In this case the UV divergence will be 
removed by replacing the bare mass $\hat m$ by the renormalized mass $m$ via
\begin{eqnarray}
\label{H8}
\hat m = m \,\Big [1 + \hat a_s\,\delta_0\,\frac{2}{\varepsilon} + \cdots 
\Big ]\,,
\end{eqnarray}
where the UV pole term is indicated by $1/\varepsilon$ with
$\varepsilon=n-4$.
If we choose for example process (\ref{H2}) together with the virtual 
corrections to the Born reaction (\ref{H1}) (see Figs. \ref{fig:4},\ref{fig:5})
the unrenormalized coefficient function takes the form
\begin{eqnarray}
\label{H9}
\hat H_{k,g}^{(2)}&=&\hat a_s^2\Bigg [\Big \{\frac{1}{\varepsilon_C}+\frac{1}{2}
\ln \left (\frac{m^2}{\mu^2}\right )\Big \}\,P_{gg}^{(0)}\otimes 
H_{k,g}^{(1)} -\beta_0\Big \{\frac{2}{\varepsilon_{UV}}
+\ln \left (\frac{m^2}{\mu^2}\right )\Big \}\,H_{k,g}^{(1)}\Bigg ]
\nonumber\\[2ex]
&& + H_{k,g}^{(2)}|_{\mu=m}\,,
\end{eqnarray}
where $H_{k,g}^{(2)}|_{\mu=m}$ is finite and
$\hat a_s$ denotes the bare coupling constant. Further we have also 
regularized the collinear divergences by n-dimensional regularization.
In order to distinguish between ultraviolet and collinear divergences
we have indicated them by $1/\varepsilon_{UV}$ and $1/\varepsilon_C$
respectively. The residues of the collinear divergences are represented
by the so called splitting functions denoted by $P_{ij}$ ($i,j=q,g$).
The origin of the collinear divergences is explained by the first diagram
in Fig. \ref{fig:5}. The propagator carrying the momentum $k_1-k_2$
behaves as
\begin{eqnarray}
\label{H10}
\frac{1}{(k_1-k_2)^2}=\frac{1}{2\,\omega_1\,\omega_2\,(1-\cos \theta)}\,,
\end{eqnarray}
which diverges for $\theta \rightarrow 0$. This propagator only shows up
if three massless particles are coupled to each other like three gluons
or when a gluon is attached to a quark line provided the quark is massless.
If the gluon is attached to a heavy quark the mass of the latter, which is 
unequal to zero, prevents that the denominator in Eq. (\ref{H10}) vanishes
when $\theta \rightarrow 0$. 
The UV-divergence in Eq. (\ref{H9}) is removed by coupling constant 
renormalization which is achieved by adding $\hat a_s\,H_{k,g}^{(1)}$
to Eq. (\ref{H9}) and replacing the bare coupling constant
$\hat a_s$ by the renormalized one represented by $a_s(\mu^2)$
\begin{eqnarray}
\label{H11}
\hat a_s = a_s(\mu^2) \,\Big [1 +  a_s(\mu^2)\,\beta_0\,\frac{2}{\varepsilon} 
+ \cdots \Big ]\,.
\end{eqnarray}
Finally one has to remove the collinear divergences. This is achieved by
mass factorization. It proceeds in a similar way as multiplicative
renormalization so that one can write
\begin{eqnarray}
\label{H12}
\hat H_{k,i}\left (\frac{1}{\varepsilon_C},Q^2,m^2\right )=
\Gamma_{ji}\left (\frac{1}{\varepsilon_C},\mu^2\right )\otimes
H_{k,j}\left (Q^2,m^2,\mu^2\right )\,,
\end{eqnarray}
where $\Gamma_{ji}$ represents the transition function which removes
all collinear divergences from the bare heavy quark coefficient functions.
Further we have introduced the notion of bare parton density $\hat f_i$ so that
the heavy quark structure function can be written as
\begin{eqnarray}
\label{H13}
F_{k,Q}(Q^2,m^2)=e_Q^2\,\hat f_i\otimes \hat H_{k,i}
\left (\frac{1}{\varepsilon_C},Q^2,m^2\right )\,.
\end{eqnarray}
Substitution of $\hat H_{k,i}$ (see Eq. (\ref{H12})) into
the expression above we can derive
\begin{eqnarray}
\label{H14}
F_{k,Q}(Q^2,m^2)=e_Q^2\,f_j(\mu^2)\otimes H_{k,j}\left (Q^2,m^2,\mu^2\right )\,,
\end{eqnarray} 
where $f_j(\mu^2)$ is the renormalized parton density defined by
\begin{eqnarray}
\label{H15}
f_j(\mu^2)=\Gamma_{ji}\left (\frac{1}{\varepsilon_C},\mu^2\right )\otimes
\hat f_i\,.
\end{eqnarray} 
In the case of the example presented in Eq. (\ref{H10}) the collinear 
divergence is removed by adding $H_{k,g}^{(1)}$ to $\hat H_{k,g}^{(2)}$
and choosing the following transition function
\begin{eqnarray}
\label{H16}
\Gamma_{gg}\left (\frac{1}{\varepsilon_C},\mu^2\right )=\delta(1-z)
+a_s\,N\,\Bigg [\frac{1}{\varepsilon_C}\,P_{gg}^{(0)} \Bigg ]\,,
\end{eqnarray}
where $N$ denotes the number of colors.
\begin{figure}[ht]
\vspace*{1mm}
\centerline{
\epsfig{file=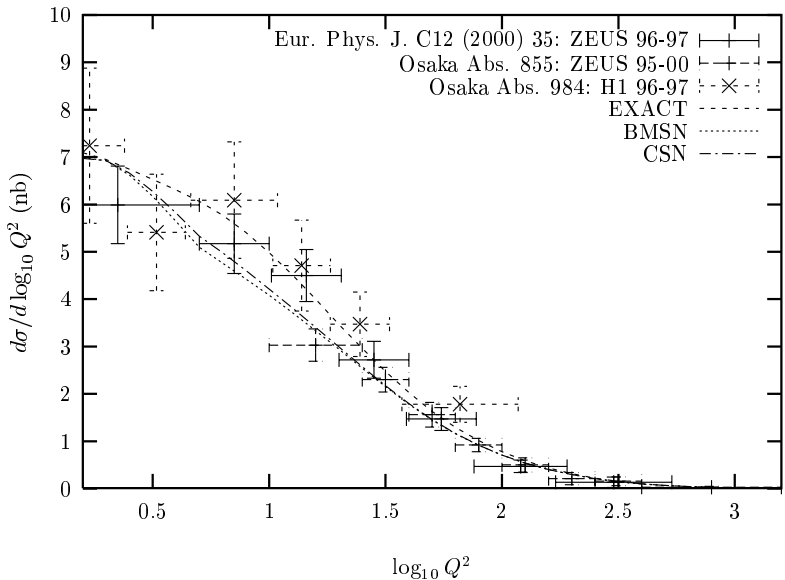,bbllx=0pt,bblly=0pt,bburx=612pt,bbury=792pt,width=11cm,%
angle=270}}
\vspace{-1.5mm}
\caption{The combined Osaka and published Zeus-data for $d\sigma/dlog_{10}Q^2$
in nb for deep inelastic production of $D^{*\pm}$-mesons. The dashed line
is the exact NLO result from the program HVQDIS which follows from
$F_{2,c}^{\rm EXACT}$. The dotted line (BMSN-scheme) and dashed-dotted line
(CSN-scheme) is based on $F_{2,c}^{\rm VFNS}$. }
\vspace{1mm}
  \label{fig:8}
\end{figure}

A comparison of the next-to-leading order (NLO) heavy quark structure function 
$F_{2,c}$ with the data for charm quark production measured at HERA reveals 
a fairly good agreement between theory and experiment. The data cover the range 
$1<Q^2<1350~{\rm GeV}^2$ and $5\times 10^{-5}<x<5.6\times 10^{-2}$.
In \cite{csh} one has made a comparison with the data obtained by the
ZEUS-collaboration \cite{breit}, \cite{zeus}. From the cross section in 
Eq. (\ref{I2}) one can derive the integrated quantities 
\begin{eqnarray}
\label{H17}
\frac{d\sigma}{dQ^2}=\int_{x_{min}}^{x_{max}}dx\,\frac{d^2\sigma}{dx\,dQ^2}\,,
\qquad
\frac{d\sigma}{dx}=\int_{Q^2_{min}}^{Q^2_{max}}dQ^2\,\frac{d^2\sigma}{dx\,dQ^2}
\,.
\end{eqnarray}
Notice that the quantities above represent $D^*_c$-meson
production rather than charm quark production. The meson appears as a 
fragmentation product of the quark and the cross sections in Eq. (\ref{H17})
are obtained by convoluting Eq. (\ref{I2}) with fragmentation functions.
Furthermore one has to impose experimental cuts on the kinematics which are 
indicated by $max$ and $min$. The results are presented in Figs. {\ref{fig:8}, 
{\ref{fig:9} which originate from \cite{csh} where one can also find
the maximal and minimal values for $x$ and $Q^2$. The figures show that there
is a good agreement between the exact NLO result (called ${\rm EXACT}$ in 
the figure) and the data except for $x \sim 10^{-3}$ where there is a small 
discrepancy. Furthermore in \cite{hasm1} 
one has also made a comparison between the program HQVDIS \cite{hasm2} 
based on the NLO computations above and the experimental
differential distributions for $D^*_c$ production where the following cross 
sections are studied.
\begin{eqnarray}
\label{H18}
\frac{d\sigma}{dp_{T,D_c}}\,, \qquad \frac{d\sigma}{d\eta_{D_c}}\,, \qquad
\frac{d\sigma}{dW}\,, 
\end{eqnarray}
where $W$ ($W^2=(p+q)^2$) is the center of mass energy of the photon-proton 
system. Also the differential distributions agree with the NLO
predictions except for the rapidity $\eta$-distribution. Here it appears that 
for $\eta_D>0$ the experimental cross section is larger than the one 
computed in NLO (see also \cite{H1}).

\begin{figure}[ht]
\vspace*{1mm}
\centerline{
\epsfig{file=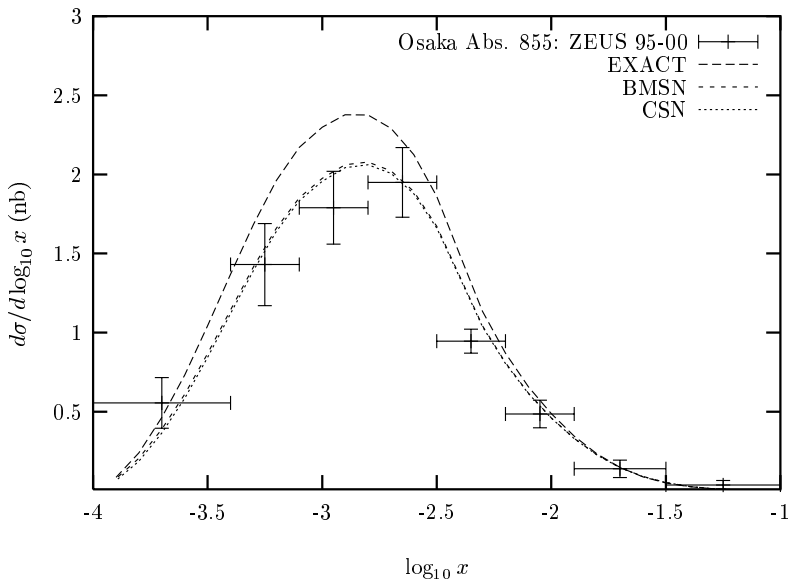,bbllx=0pt,bblly=0pt,bburx=612pt,bbury=792pt,width=11cm,%
angle=270}}
\vspace{-1.5mm}
\caption{The combined Osaka and published Zeus-data for $d\sigma/dlog_{10}x$ 
in nb for deep inelastic production of $D^{*\pm}$-mesons. The dashed line
is the exact NLO result from the program HVQDIS which follows from
$F_{2,c}^{\rm EXACT}$. The dotted line (BMSN-scheme) and dashed-dotted line 
(CSN-scheme) is based on $F_{2,c}^{\rm VFNS}$. }
\label{fig:9}
\vspace{1mm}
\end{figure}

Summarizing our findings for the exact NLO calculations we conclude
\begin{itemize}
\item[1.]
There exist a fairly good agreement between the data and the NLO calculations.
\item[2.]
The theoretical curves are insensitive to the choice of the 
renormalization/factorization scale $\mu$ occurring in the parton densities
and the coefficient functions \cite{grs}.
\item[3.]
The theoretical curves are sensitive to the choice of the charm mass $m_c$
which is in the range $1.3 <m_c<1.7~{\rm GeV/c^2}$.
\item[4.]
Processes with a gluon in the initial state (Eqs. (\ref{H1}),(\ref{H2}))
constitute the bulk of the
contribution to the heavy quark structure function at $x<10^{-2}$. Hence
they are an excellent probe to measure the gluon density $f_g(z,\mu^2)$. 
\end{itemize}
\begin{center}
\section{Asymptotic Heavy Quark Structure Functions}
\end{center}

In this section we want to study the heavy quark coefficient functions
$H_{k.i}$, $L_{k,i}$ in the asymptotic region $Q^2 \gg m^2$. In this region
they have the following form
\begin{eqnarray}
\label{A1}
H_{k,i}^{\rm ASYMP}(z,Q^2,m^2,\mu^2)\sim \sum_{l=1}^{\infty}a_s^l
\sum_{n+m\le l} a_{nm}(z)\,
\ln^n\left (\frac{\mu^2}{m^2}\right )\,\ln^m\left (\frac{Q^2}{m^2}\right )\,.
\end{eqnarray}
A similar expression exists for $L_{k,i}^{\rm ASYMP}$. An example is the
asymptotic expression for the Born reaction in Eq. (\ref{H1}) 
(Fig. \ref{fig:3}) which can be written as
\begin{eqnarray}
\label{A2}
H_{2,g}^{\rm ASYMP}(z,Q^2,m^2)=a_s\Bigg [\frac{1}{2}P_{qg}(z)
\ln\left (\frac{Q^2}{m^2}\right ) + a_{Qg}(z)+c_{2,g}(z)\Bigg ]\,.
\end{eqnarray}
The origin of this asymptotic behaviour can be attributed to the property
that in the limit $m\rightarrow 0$ the heavy quark coefficient functions
become collinear divergent which is revealed by the logarithmic singularities
$\ln Q^2/m^2$ and $\ln \mu^2/m^2$. The reason why this behaviour is of interest
can be summarized as follows
\begin{itemize}
\item[1.]
The results obtained for the exact coefficient functions in the previous section
are semi-analytic. In \cite{lrsn} one has obtained exact results for the virtual
corrections to process (\ref{H1}) (Fig. \ref{fig:4}) but for the reactions
with three particles in the final state like Eqs. (\ref{H2}),(\ref{H3}),
(\ref{H6}) a full analytical expression could only be presented for the 
Compton process in Eq. (\ref{H6}) (see \cite{bmsmn}). In the other cases only
the integration over the angles could be carried out but the integration over
the final state energies are so tedious that they have to be done in a numerical
way. However the latter integration becomes more amenable when $m^2\ll Q^2$
so that terms of the order $m^2/Q^2$ can be neglected. Therefore an analytical 
result for the asymptotic heavy quark coefficient functions provides us with 
a check of the exact expressions computed for arbitrary $m$.
\item[2.]
The asymptotic coefficient functions play an important role in the derivation
of the variable flavor number scheme \cite{acot}, \cite{bmsn} discussed at the 
end of this section.
This scheme is only useful if the following questions are answered. They are:
\begin{itemize}
\item[a.]
Do the logarithmic terms of the type $\ln^n Q^2/m^2$, occurring in the 
coefficient functions, really dominate the heavy quark structure functions
or the heavy quark cross sections?
\item[b.]
Are the logarithmic terms $\ln^n Q^2/m^2$ so large that they bedevil the
perturbation series so that they have to be re-summed?
\end{itemize} 
\end{itemize}
There are two ways to compute the asymptotic heavy quark coefficient functions.
\begin{itemize}
\item[1.]
One can follow the procedure for the derivation of the exact coefficient
functions in \cite{lrsn} but since the mass $m$ can be neglected one can
now carry out the additional integrations over the energies of
the final state particles in an analytical way.
\item[2.]
One can use the operator product expansion (OPE) techniques which however
are only applicable for inclusive quantities like the structure functions
$F_{k,Q}(x,Q^2,m^2)$.
\end{itemize}
In \cite{bmsmn} one has adopted the latter approach which will be outlined 
below.
In the derivation of the cross section in Eq. (\ref{I3}) one encounters
the hadronic tensor leading to the definition of the structure functions.
It is defined by
\begin{eqnarray}
\label{A3}
W_{\mu\nu}(p,q)&=&\frac{1}{4\pi}\,\int d^4z\,e^{iq\cdot z}\,\langle P(p)
|\Big [J_{\mu}(z),J_{\nu}(0)\Big ]|P(p)\rangle 
\nonumber\\[2ex]
&=&\Big (p_{\mu}p_{\nu}-
\frac{p\cdot q}{q^2}(p_{\mu}q_{\nu}+p_{\nu}q_{\mu})+g_{\mu\nu}
\frac{(p\cdot q)^2}{q^2}\Big )\frac{F_2(x,Q^2,M^2)}{p\cdot q}
\nonumber\\[2ex]
&+&\Big (g_{\mu\nu}-\frac{q_{\mu}q_{\nu}}{q^2}\Big )\frac{F_L(x,Q^2,M^2)}{2~x}
\,,
\nonumber\\[2ex]
&& p^2=M^2\,, \qquad q^2=-Q^2\,, \qquad x=\frac{Q^2}{2p\cdot q}\,.
\end{eqnarray}
In the limit $Q^2\gg M^2$ the current-current correlation function appearing
in the integrand of Eq. (\ref{A3}) is dominated by the light cone. Hence
one can make an operator product expansion near the light cone and write
\begin{eqnarray}
\label{A4}
\Big [J(z),J(0)\Big ]
\mathop{\sim}\limits_{\vphantom{\frac{A}{A}} z^2 \rightarrow 0}
\sum_i\sum_m c_i^{(m)}(z^2\mu^2)\,z_{\mu_1}\cdots z_{\mu_m}\,
O_i^{\mu_1\cdots \mu_m}(0,\mu^2)\,,
\end{eqnarray}
where for convenience we have dropped the Lorentz indices of the currents.
Here $c_i^{(m)}$ denote the coefficient functions which are distributions
and $O_i^{\mu_1\cdots \mu_m}$ are local operators. Both are renormalized
which is indicated by the renormalization scale $\mu$. When dropping all
terms of the order $M^2/Q^2$ we can limit ourselves to leading twist
operators. In QCD they are given by
\begin{eqnarray}
\label{A5}
&&\mbox{non-singlet operators}
\nonumber\\[2ex]
&& O_{q,r}^{\mu_1\cdots \mu_m}(z)=\frac{1}{2}\,i^{m-1}\,{\cal S}
\Big [\bar \psi(z)\,\gamma^{\mu_1}\,D^{\mu_2}\cdots D^{\mu_m}\,
\frac{\lambda_r}{2}\,\psi(z)\Big ]+\mbox{trace terms}\,,
\nonumber\\[2ex]
&&\mbox{singlet operators}
\nonumber\\[2ex]
&& O_q^{\mu_1\cdots \mu_m}(z)=\frac{1}{2}\,i^{m-1}\,{\cal S}
\Big [\bar \psi(z)\,\gamma^{\mu_1}\,D^{\mu_2}\cdots D^{\mu_m}\,\psi(z)\Big ]
+\mbox{trace terms}\,,
\nonumber\\[2ex]
&& O_g^{\mu_1\cdots \mu_m}(z)=\frac{1}{2}\,i^{m-2}\,{\cal S}\Big [ 
F_{\alpha}^{a,\mu_1}(z)\,D^{\mu_2} \cdots D^{\mu_{m-1}}\,
F_{\alpha}^{a,\alpha\mu_m}(z)]+\mbox{trace terms}\,.
\end{eqnarray}
The symbol ${\cal S}$ indicates that one has to symmetrize over all Lorentz
indices. The covariant derivative is given by $D^{\mu}=\partial^{\mu}
-i\,g\,T_a\,A_a^{\mu}$ and $\lambda_r$ is the generator of the flavor
group $SU(n_f)_F$. Insertion of the OPE (Eq. (\ref{A4})) into the hadronic
tensor of Eq. (\ref{A3}) leads to the following result
\begin{eqnarray}
\label{A6}
F_k^{(m)}(Q^2,M^2)\equiv \int_0^1dx\,x^{m-1}\,F(x,Q^2)=\sum_{i=q,g}A_i^{(m)}
\left (\frac{\mu^2}{M^2}\right )
\,{\cal C}_i^{(m)}\left (\frac{Q^2}{\mu^2}\right )\,,
\end{eqnarray}
where the operator matrix element and the coefficient function are defined by
\begin{eqnarray}
\label{A7}
A_i^{(m)}\left (\frac{\mu^2}{M^2}\right )=\langle P(p)|O_i^m(0,\mu^2)|P(p)
\rangle\,, \quad {\cal C}_i^{(m)}\left (\frac{Q^2}{\mu^2}\right )=
\int d^4z\,e^{iq\cdot z}\,c_i^{(m)}(z^2\mu^2)\,.
\end{eqnarray}
The OPE techniques can also be applied when the the proton state $|P(p)\rangle$
in Eq. (\ref{A3}) is replaced by a light quark state $|q(p)\rangle$ or
a gluon state $|g(p)\rangle$. However when the proton is replaced by massless 
quarks and gluons the external momentum satisfies the relation $p^2=0$ so that 
the partonic structure functions and the partonic operator 
matrix elements become collinearly divergent.
One can show (see \cite{bmsn}) that instead of Eq. (\ref{A6}) one obtains more 
complicate expressions which are given by
\begin{eqnarray}
\label{A8}
&&\hat {\cal F}_{k,q}^{\rm NS}\left (\frac{Q^2}{\mu^2},
\frac{1}{\varepsilon_C}\right )
+\hat L_{k,q}^{\rm ASYMP}\left (\frac{Q^2}{m^2},\frac{m^2}{\mu^2}
,\frac{1}{\varepsilon_C}\right )=\hat A_{qq}^{\rm NS}\left (\frac{m^2}{\mu^2},
\frac{1}{\varepsilon_C}\right )\otimes {\cal C}_{k,q}^{\rm NS}
\left (\frac{Q^2}{\mu^2}\right )\,,
\nonumber\\[2ex]
&&\hat {\cal F}_{k,i}^{\rm S}\left (\frac{Q^2}{\mu^2},\frac{1}{\varepsilon_C}
\right )
+\hat L_{k,i}^{\rm ASYMP}\left (\frac{Q^2}{m^2},\frac{m^2}{\mu^2}
,\frac{1}{\varepsilon_C}\right )
+ \hat H_{k,i}^{\rm ASYMP}\left (\frac{Q^2}{m^2},\frac{m^2}{\mu^2},
\frac{1}{\varepsilon_C} \right )
\nonumber\\[2ex]
&&=\hat A_{ji}^{\rm S}\left (\frac{m^2}{\mu^2},\frac{1}{\varepsilon_C}\right )
\otimes {\cal C}_{k,j}^{\rm NS} \left (\frac{Q^2}{\mu^2}\right )\,,
\nonumber\\[2ex]
&& i,j=q,g\,.
\end{eqnarray}
Here $\hat {\cal F}_{k,i}$ represent the partonic structure functions which
are given by Feynman graphs containing massless particles only and therefore
contain collinear singularities indicated by $\varepsilon_C$. These
singularities also appear in the asymptotic heavy quark coefficient functions
$\hat H_{k,i}^{\rm ASYMP}$, $\hat L_{k,i}^{\rm ASYMP}$ which are determined in 
the asymptotic regime $Q^2\gg m^2$ before mass factorization is carried out. 
These collinear divergences can be traced back to the massless quarks and
gluons appearing in Figs. \ref{fig:3}-\ref{fig:7}. Finally $\hat A_{ji}$ 
represent the operator matrix elements on the partonic level and are defined
by
\begin{eqnarray}
\label{A9}
\hat A_{ji}\left (\frac{m^2}{\mu^2},\frac{1}{\varepsilon_C}\right )
=\langle i |O_j(\mu^2,0)| i \rangle\,, \qquad i=q,g\,,\qquad j=q,g,Q\,, 
\qquad p_i^2=0\,.
\end{eqnarray}
The operator matrix elements which depend on the heavy flavor mass consist of 
two classes. The first class is given by heavy quark operators sandwiched
between gluon or light quark states. The second class contains light
quark or gluon operators which contain a heavy flavor loop. Examples of the 
first class together with the corresponding process are given in Figs.
\ref{fig:10} and \ref{fig:11}. An example of the second class is shown in 
\ref{fig:12}.
\begin{figure}
\begin{center}
  \begin{picture}(120,80)
  \Gluon(7,0)(30,30){3}{5}
  \ArrowLine(60,80)(30,30)
  \ArrowLine(90,30)(60,80)
  \ArrowLine(30,30)(90,30)
  \Gluon(90,30)(113,0){3}{5}
   \Text(60,85)[t]{$\otimes$}
   \Text(60,25)[t]{$Q$}
   \Text(35,60)[t]{$Q$}
   \Text(85,60)[t]{$Q$}
   \Text(4,15)[t]{$g$}
   \Text(116,15)[t]{$g$}
  \end{picture}
\hspace*{1cm}
  \begin{picture}(130,80)
  \Gluon(0,0)(30,30){3}{5}
  \ArrowLine(60,30)(30,30)
  \ArrowLine(30,30)(30,65)
  \ArrowLine(30,65)(60,65)
  \Photon(0,85)(30,65){3}{5}
  \DashLine(65,85)(65,0){3}
  \Gluon(130,0)(100,30){3}{5}
  \ArrowLine(70,30)(100,30)
  \ArrowLine(100,30)(100,65)
  \ArrowLine(100,65)(70,65)
  \Photon(130,85)(100,65){3}{5}
   \Text(0,15)[t]{$g$}
   \Text(130,15)[t]{$g$}
   \Text(22,52)[t]{$Q$}
   \Text(110,52)[t]{$Q$}
   \Text(10,95)[t]{$\gamma^*$}  
   \Text(120,95)[t]{$\gamma^*$}
  \end{picture}
\vspace*{5mm}
  \caption{The operator matrix element $A_{Qg}^{(1)}$ and the corresponding
Feynman graph for the process $\gamma^* + g \rightarrow Q + \bar Q$
         ($H_{k,g}^{(1)}$). }
  \label{fig:10}
\end{center}
\end{figure}
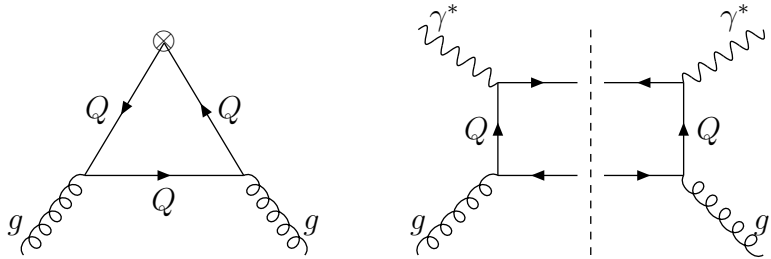
The light partonic coefficient functions defined by ${\cal C}_{k,j}$
are derived from the light partonic structure functions via mass factorization
\begin{eqnarray}
\label{A10}
\hat {\cal F}_{k,i}\left (\frac{Q^2}{\mu^2},\frac{1}{\varepsilon_C}\right )
=\Gamma_{ji}\left (\frac{1}{\varepsilon_C},\mu^2\right )\otimes
{\cal C}_{k,j}\left (\frac{Q^2}{\mu^2}\right )\,,
\end{eqnarray}
where $\Gamma_{ji}$ denote the transition functions which are discussed
below Eq. (\ref{H12}).
\begin{figure}
\begin{center}
  \begin{picture}(130,130)
  \Gluon(0,0)(30,30){3}{5}
  \ArrowLine(65,110)(30,70)
  \ArrowLine(30,70)(100,70)
  \ArrowLine(100,70)(65,110)
  \Gluon(30,30)(100,30){3}{8}
  \Gluon(30,30)(30,70){3}{5}
  \Gluon(100,70)(100,30){3}{5}
  \Gluon(130,0)(100,30){3}{5}
   \Text(65,115)[t]{$\otimes$}
  \end{picture}
\hspace*{1cm}
  \begin{picture}(130,130)
  \Gluon(0,0)(30,20){3}{5}
  \Gluon(30,20)(60,20){3}{4}
  \ArrowLine(60,80)(30,80)
  \ArrowLine(30,80)(30,50)
  \ArrowLine(30,50)(60,50)
  \Gluon(30,20)(30,50){3}{4}
  \Photon(0,100)(30,80){3}{5}
  \DashLine(65,100)(65,0){3}
  \Gluon(100,20)(130,0){3}{5}
  \Gluon(100,20)(70,20){3}{4}
  \ArrowLine(70,80)(100,80)
  \ArrowLine(100,80)(100,50)
  \ArrowLine(100,50)(70,50)
  \Gluon(100,20)(100,50){3}{4}
  \Photon(130,100)(100,80){3}{5}
  \end{picture}
\vspace*{5mm}
  \caption{The operator matrix element $A_{Qg}^{(2)}$ and the corresponding
Feynman graph for the process $\gamma^* + g \rightarrow Q + \bar Q + g$
($H_{k,g}^{(2)}$). }
  \label{fig:11}
\end{center}
\end{figure}
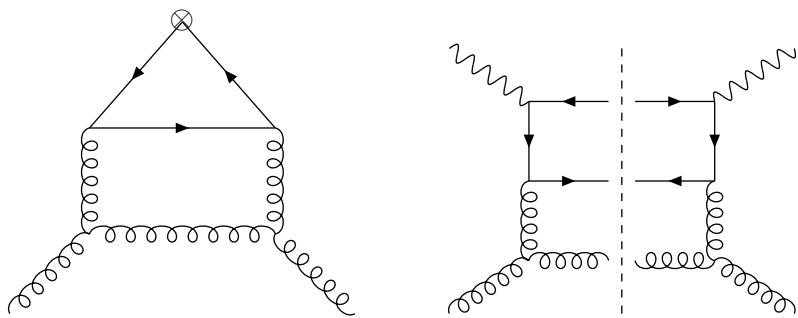
The quantities $\hat {\cal F}_{k,i}$ and ${\cal C}_{k,i}$ have been calculated 
up to second order in $\alpha_s$ in \cite{zn}. The operator matrix elements 
$\hat A_{ij}$
are also known up to second order and the calculation of them is presented
in \cite{bmsmn}, \cite{bmsn}. Like in the case of the coefficient functions 
n-dimensional regularization has been used to regularize the ultraviolet and 
collinear 
divergences and one has chosen the same renormalization conditions for the mass 
and the strong coupling constant. Hence from Eq. (\ref{A8}) one infers the 
asymptotic heavy quark coefficient functions $\hat H_{k,i}^{\rm ASYMP}$, 
$\hat L_{k,i}^{\rm ASYMP}$ up to the same order (see \cite{bmsn}). 
Finally one can remove the remaining collinear divergences via the same 
mass factorization as is done for the exact heavy quark coefficient functions 
in Eq. (\ref{H12}).
\begin{figure}
\begin{center}
  \begin{picture}(130,120)
    \DashArrowLine(10,10)(25,40){5}
    \DashArrowLine(25,40)(65,120){5}
    \DashArrowLine(65,120)(105,40){5}
    \DashArrowLine(105,40)(120,10){5}
   \ArrowArc(65,40)(15,0,180)
   \ArrowArc(65,40)(15,180,360)
  \Gluon(25,40)(50,40){3}{4}
  \Gluon(80,40)(105,40){3}{4}
   \Text(65,125)[t]{$\otimes$}
   \Text(0,10)[t]{$q$}
   \Text(130,10)[t]{$q$}
   \Text(38,84)[t]{$q$}
   \Text(92,84)[t]{$q$}
   \Text(65,72)[t]{$\bar Q$}
   \Text(65,22)[t]{$Q$}
  \end{picture}
\hspace*{1cm}
  \begin{picture}(130,130)
  \Photon(0,120)(25,100){3}{5}
  \Photon(130,120)(105,100){3}{5}
    \DashArrowLine(0,10)(25,40){5}
    \DashArrowLine(25,40)(25,100){5}
    \DashArrowLine(25,100)(60,100){5}
    \DashArrowLine(130,10)(105,40){5}
    \DashArrowLine(105,40)(105,100){5}
    \DashArrowLine(105,100)(70,100){5}
   \ArrowArc(60,40)(15,90,180)
   \ArrowArc(60,40)(15,180,270)
   \ArrowArcn(70,40)(15,90,0)
   \ArrowArcn(70,40)(15,360,270)
  \Gluon(25,40)(45,40){3}{3}
  \Gluon(85,40)(105,40){3}{3}
  \DashLine(65,120)(65,10){3}
   \Text(12,130)[t]{$\gamma^*$}  
   \Text(121,130)[t]{$\gamma^*$}  
   \Text(10,10)[t]{$q$}
   \Text(126,10)[t]{$q$}
   \Text(112,71)[t]{$q$}
   \Text(18,71)[t]{$q$}
   \Text(55,70)[t]{$\bar Q$}
   \Text(55,24)[t]{$Q$}
  \end{picture}
  \caption{The operator matrix element $A_{qq}^{\rm NS,(2)}$ and the 
   corresponding Feynman graph for the process 
   $\gamma^* + q \rightarrow Q + \bar Q + q$ ($L_{k,q}^{\rm NS,(2)}$). }
  \label{fig:12}
\end{center}
\end{figure}
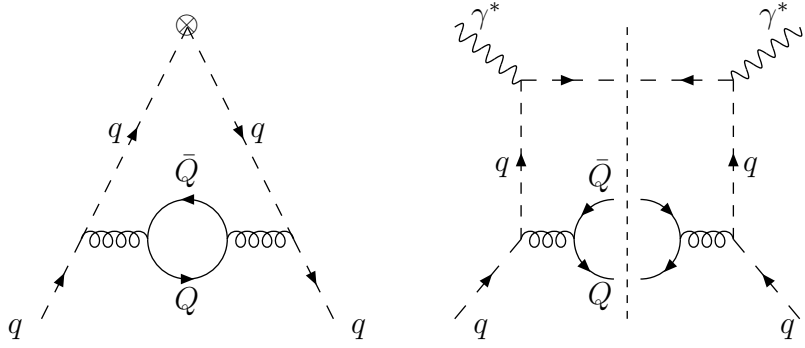
After this outline of the calculation of the asymptotic heavy quark
coefficient function one might ask the question whether it is not easier
to compute them in a more direct way. The main problem of radiative corrections 
is the computation of the phase space integrals in particular if one
has massive particles in the final state even if one takes $m^2\ll Q^2$.
Since this work was already done for the light partonic structure functions
$\hat {\cal F}_{k,i}$ in \cite{zn} it was not needed to repeat this procedure 
anymore. On the contrary it is much easier to compute two-loop operator matrix
elements because of the zero momentum flowing into the operator vertex
indicated by the symbol $\otimes$ in Figs. \ref{fig:10}-\ref{fig:12}.
\begin{figure}[ht]
\vspace*{1mm}
\centerline{
\epsfig{file=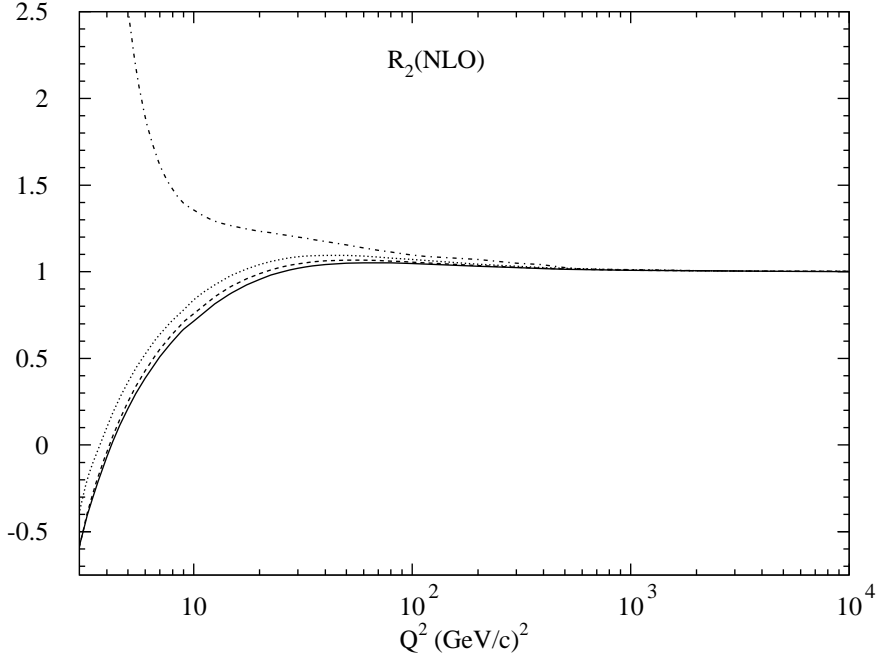,bbllx=0pt,bblly=0pt,bburx=612pt,bbury=792pt,width=11cm,%
angle=270}}
\vspace{-1.5mm}
\caption{$R_{2,c}$ in NLO (Eq. (\ref{A11})) plotted as function of $Q^2$ at 
fixed $x$; $x=10^{-1}$ (dashed-dotted line), $x=10^{-2}$ (dotted line),
$x=10^{-3}$ (dashed line), $x=10^{-4}$ (solid line) .}
\label{fig:13}
\vspace{1mm}
\end{figure}
The difference between the exact and asymptotic coefficient functions
can be attributed to the power corrections of the type $(m^2/Q^2)^l$
,with $l\ge 1$, which occur in the former but are absent in the latter. The 
asymptotic coefficient functions only contain the logarithms $\ln^m Q^2/m^2$
and $\ln^n \mu^2/m^2$ and terms which survive in the limit 
$Q^2 \rightarrow \infty$. We will now first answer the question whether
these logarithmic terms dominate the heavy quark structure function 
$F_{k,Q}(x,Q^2,m^2)$. For that purpose we have studied the structure
functions for charm production i.e. $Q=c$ in \cite{bmsn}, \cite{laen}. 
Here one has computed the ratio
\begin{eqnarray}
\label{A11}
R_{k,c}=\frac{F_{k,c}^{\rm ASYMP}}{F_{k,c}^{\rm EXACT}}\,,
\end{eqnarray}
where $F_{k,c}^{\rm EXACT}$ and $F_{k,c}^{\rm ASYMP}$ are represented in
the three flavor number scheme
\begin{eqnarray}
\label{A12}
&& F_{k,c}^{\rm EXACT}=
\nonumber\\[2ex]
&& \frac{4}{9}\,\sum_{i=q,g}f_i^{\rm S}(3,\mu^2)\otimes
H_{k,i}^{\rm EXACT}+\sum_{j=u,d,s}e_j^2\,(f_j(3,\mu^2)+f_{\bar j}(3,\mu^2))
\otimes L_{k,q}^{\rm EXACT}\,,
\nonumber\\
\end{eqnarray}
\newpage
\begin{eqnarray}
\label{A13}
&& F_{k,c}^{\rm ASYMP}=
\nonumber\\[2ex]
&& \frac{4}{9}\,\sum_{i=q,g}f_i^{\rm S}(3,\mu^2)\otimes
H_{k,i}^{\rm ASYMP}+\sum_{j=u,d,s}e_j^2\,(f_j(3,\mu^2)+f_{\bar j}(3,\mu^2))
\otimes L_{k,q}^{\rm ASYMP}\,.
\nonumber\\
\end{eqnarray}
In Fig. \ref{fig:13} we have plotted $R_{2,c}$. Here one observes that this 
quantity becomes
very close to one for $x<10^{-2}$ and $Q^2>20~{\rm GeV}^2$ which belongs to
the region explored by the experiments carried out at HERA. This shows
that the logarithms mentioned above dominate the structure function except
in the threshold region given by $x \sim 1$ and small $Q^2$ which is
characteristic of the EMC experiment \cite{aubert}.

In order to answer the second question whether these logarithmic terms vitiate
the perturbation series one has to resum them in all orders of perturbation
theory and show that the resummed structure function differs from the one
which is computed exactly in fixed order of perturbation theory. This 
resummation procedure is provided by the variable flavor number scheme
(VFNS) \cite{acot}. An example of a resummed structure function has been 
derived in \cite{bmsn}. In the case of charm quark production one obtains
\begin{eqnarray}
\label{A14}
F_{k,c}^{\rm VFNS}= \sum_{j=u,d,s,c}e_j^2\,\Bigg [(f_j(4,\mu^2)
+f_{\bar j}(4,\mu^2))
\otimes {\cal C}_{k,j}^{\rm VFNS}+f_g(4,\mu^2)\otimes{\cal C}_{k,g}^{\rm VFNS}
\Bigg ]\,,
\end{eqnarray}
with the conditions
\begin{eqnarray}
\label{A15}
\mathop{{\rm lim}}\limits_{\vphantom{\frac{A}{A}} m^2\rightarrow 0}
{\cal C}_{k,i}^{\rm VFNS}\left (\frac{Q^2}{m^2},\frac{\mu^2}{m^2}\right )
={\cal C}_{k,i}\left (\frac{Q^2}{\mu^2}\right )\,,
\end{eqnarray}
\begin{eqnarray}
\label{A16}
\mathop{{\rm lim}}\limits_{\vphantom{\frac{A}{A}} s \rightarrow  4m^2}
F_{k,c}^{\rm VFNS}(x,Q^2,m^2)
=F_{k,c}^{\rm EXACT}(x,Q^2,m^2)\,, \qquad s=\frac{1-x}{x}\,Q^2\,.
\end{eqnarray}
The consequence of condition (\ref{A15}) is that in the asymptotic regime
$Q^2\gg m^2$, $F_{k,c}^{\rm VFNS}$ turns into the structure function
represented in the four flavor number scheme which
contains the contribution of light flavors only including the charm quark. 
The mass singular logarithms, occurring in the heavy quark coefficient 
functions in the three flavor number scheme (see Eqs. (\ref{H4}), (\ref{H7})),
are shifted to the parton densities defined in the four flavor number scheme 
appearing in Eq. (\ref{A14}). The most conspicuous feature is the 
appearance of the charm quark density which is absent in a three flavor number 
scheme but shows up in the four flavor number scheme. It is given by 
\begin{eqnarray}
\label{A17}
f_c(4,\mu^2)+f_{\bar c}(4,\mu^2)=f_q^{\rm S}(3,\mu^2)\otimes A_{Qq}^{\rm S}
\left (\frac{\mu^2}{m^2}\right )+f_g(3,\mu^2)\otimes A_{Qg}^{\rm S}
\left (\frac{\mu^2}{m^2}\right )\,.
\end{eqnarray}
The operator matrix elements satisfy renormalization group equations which 
enable us to resum all logarithmic terms of the type $\ln \mu^2/m^2$.
In order to get the boundary condition (\ref{A16}) one needs matching 
conditions for which one can make various choices (see e.g. \cite{acot},
\cite{bmsn}, \cite{thro}). Two of them are proposed in \cite{bmsn},\cite{neer2}
(BMSN scheme) and in \cite{csn} (CSN scheme). In the former one equates
\begin{eqnarray}
\label{A18}
{\cal C}_{k,c}^{\rm VFNS}\left (\frac{Q^2}{m^2},\frac{\mu^2}{m^2}\right )
={\cal C}_{k,q}\left (\frac{Q^2}{\mu^2}\right )\,,\qquad q=u,d,s\,.
\end{eqnarray}
Both schemes have been calculated up to next-to-next-to-leading order (NNLO) 
and are compared in \cite{csh} with $F_{k,c}^{\rm EXACT}$ (NLO) in 
Eq. (\ref{A12}). The results are shown in Figs. \ref{fig:8}, \ref{fig:9} 
from which one infers that there is hardly any difference between the two 
schemes representing VFNS. This shows that one can neglect the power 
contributions ${\cal O}(m^2/Q^2)$ in ${\cal C}_{k,c}^{\rm VFNS}$
which are absent in ${\cal C}_{k,q}$. Also the difference
between the two versions of VFNS on one hand and the exact NLO approach 
on the other hand is hardly noticeable except in Fig. \ref{fig:9} where
in the vicinity of $x=10^{-3}$ 
it seems that the data are better described by the BMSN and CSN schemes
than by the exact NLO result. The main conclusion that one can draw from
these figures is that the resummation effect is very small which means
that the so called large logarithms of the type $\ln Q^2/m^2$ and also
$\ln \mu^2/m^2$, when $\mu^2 \sim Q^2$, do not vitiate the perturbation series.

Summarizing our results we conclude
\begin{itemize}
\item[1.]
The past ten years have shown much progress in the computation of
higher order corrections to heavy flavor production. In particular
the results obtained in electro-production agree well with the data
obtained by the experiments carried out at HERA.
\item[2.]
The asymptotic heavy quark coefficient functions can be calculated using
operator product expansion techniques. The results obtained for the
operator matrix elements can be also used for processes where the light
cone does not dominate the reaction which e.g. holds for 
$e^+~e^- \rightarrow \mu^+~\mu^-$ \cite{bnb}.
\item[3.]
The heavy quark structure function is dominated by the logarithmic terms
$\ln Q^2/m^2$ and $\ln \mu^2/m^2$ provided $x$ and $Q^2$ are chosen in such
a way that they are outside the threshold region of the production process
i.e.\\
$s=(1-x)Q^2/x \gg 4m^2$.
\item[4.]
In spite of the fact that the logarithms above dominate the structure function
they do not bedevil the convergence of the perturbation series so that a
resummation is in principle not necessary. Therefore one can use fixed order
(exact) perturbation theory which is simple to apply and to interpret
in particular if one studies the differential distributions
presented in Eqs. (\ref{I4}), (\ref{H18}).
\end{itemize}

\end{document}